\documentclass[pre,amsmath,twocolumn,floatfix,showpacs]{revtex4}

\usepackage{graphicx}

\usepackage{amssymb}

\begin{document}

\title{Reorientation dynamics in thin glassy films}
\author{E. Cecchetto, D. Moroni, B. J{\'e}r{\^o}me}
\affiliation{Department of Chemical Engineering, University of Amsterdam Nieuwe Achtergracht 166, 1018WV Amsterdam, The Netherlands}
\date{\today}

\begin{abstract}
We present a study of orientational relaxation dynamics in thin films of a low-molecular-weight glass-former as a function of temperature and film thickness. 
The relaxation is probed by second-harmonic generation after release of a poling electric field. 
From the measured decays of the second-harmonic signal and their fitting with a stretched exponential, we can determine the distribution of relaxation times in the system. 
As temperature decreases from above the glass transition, we observe that the width of the distribution first increases under confinement, but that deeper in the glassy state, confinement has no effect anymore on the dynamics.
\end{abstract}


\maketitle
\cleardoublepage
\section{INTRODUCTION}
In the liquid phase of glass formers, the viscosity and any sort of relaxation time increase enormously upon cooling. 
Relaxation times eventually become longer than the experimental time scale, which is essentially the time during which the system is allowed to equilibrate after a change of an external parameter or constraint.
As a consequence, no changes are observable anymore in the system and the structural configuration of the liquid appears to freeze below the so-called glass transition temperature $T_g$.
A deep understanding of the nature of the glass transition it is still lacking and reaching it is a challenging goal. 
Many theories have been proposed \cite{Gotze,80,320,19,342} and revisited, but all fail to explain the whole collection of typical features of the glass transition. 
Somehow each theory applies to a limited range of conditions, either thermodynamic or kinetic.

One of the widely used theories of the glass transition is that of Adam and Gibbs \cite{69}, which is based on the concept of cooperatively rearranging regions (CRRs). 
According to this model, as the temperature of a liquid decreases toward the glass transition, the motion of the molecules is no longer independent, but the dynamics becomes correlated over a certain number of molecules. 
The spatial extent of this correlation is generally defined as the cooperative length $\xi$. 
Parallel to the development of theories based on the idea of cooperative motion, other models focused on explaining the fact that the dominant $\alpha$-relaxation is highly non-exponential. 
This non-exponentiality can be related to the superposition of relaxation processes with different relaxation times associated with different regions in the material, the so-called dynamic heterogeneities. 
The concept is indeed related to the CRRs, which can be considered as fast relaxing regions surrounded by slow or immobile molecules \cite{334}. 
So the length scale of heterogeneities and the cooperative length should essentially be the same. 
Both simulation and experiments have been used to provide evidence for the existence of such a length scale and to evaluate it (see i.e \cite{334,340,61,73,331,335}).
The obtained values at $T_g$ or just above it, are of the order of a few nanometers, both for low-molecular weight and polymeric systems. 
Theories predict that this length should increase as temperature decreases. 

\begin{figure}[t!]
\begin{center}
\noindent
{\includegraphics[width=8cm,keepaspectratio]{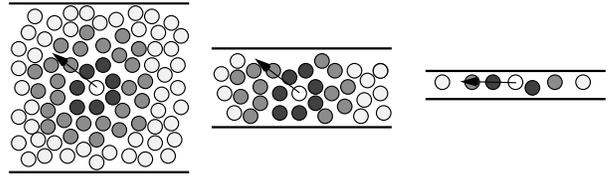}}
\caption{\small Schematic representation of the effect of confinement on cooperative motion: as the size of the system decreases, cooperatively rearranging regions (the radius of which is indicated by the arrow) gets modified, hence also the dynamics of the system.}
\label{confining}
\end{center}
\end{figure}
The concept of cooperativity suggests that the dynamics of glass-formers should be modified if the size of the system becomes comparable to the cooperative length $\xi$. 
In this size range, it is the size of the system that determines the size of cooperatively rearranging regions (Fig.~\ref{confining}) and therefore the dynamic behavior of the system. 
So one expects modifications of the dynamics, namely of the distribution of relaxation time and the glass transition temperature, in glassy systems with sizes in the mesoscopic range. 
Taking into account only pure geometrical effects (without interactions with walls) and assuming the presence of cooperatively rearranging regions of size $\xi$, we can expect the following changes in the dynamic behavior of glass-formers when they are confined in thin films \cite{267}. 
For thicknesses rather large with respect to $\xi$, confinement only affects molecules in the vicinity of the surfaces (Fig.~\ref{confining}). 
Surface molecules belong to cooperatively rearranging regions that are smaller than the ones present in bulk. 
Therefore their relaxation should be faster than that of molecules in the middle of the system. 
As the size of the system decreases, the proportion of molecules affected by the surfaces increases. 
As a consequence, the distribution of relaxation times in the system should broaden toward the short time scales leading to a decrease of the mean relaxation time. 
When the film thickness becomes comparable with $\xi$, all molecules are affected by confinement and the size of the cooperatively rearranging regions is determined by the size of the system. 
Decreasing the size even more down to a molecular size will make the system homogeneous again, with all molecules having essentially the same environment. 
However, the relaxation time will be smaller than the value in bulk, because the motion of the molecules becomes essentially individual again.

\begin{figure}[t!]
\begin{center}
\noindent
{\includegraphics[width=6.0cm,keepaspectratio]{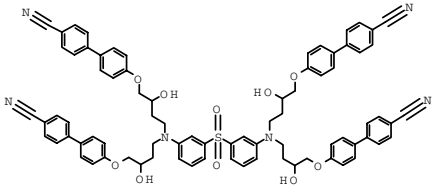}}
\caption{\small Chemical structure of the glass-former.}
\label{molecules}
\end{center}
\end{figure}
Research on the effect of confinement on glass dynamics has concentrated on two types of systems: low-molecular-weight glass-formers in porous media and polymers in thin films. 
In low-molecular-weight glass-formers, as measured by dielectric spectroscopy, the characteristic time corresponding to the $\alpha$ orientational relaxation is generally decreasing and the distribution of relaxation times is broadening as the size of the system decreases \cite{34,10,370,45}. 
From the effects of size on the dynamics, it is possible to make an estimate of the cooperative length $\xi$. 
The value of $\xi$ for salol in the vicinity of the calorimetric glass transition temperature was estimated to be $\ge$ 70 \AA\ \cite{10}. Similar values of $\xi$ at $T_{g}$ were obtained for propylene glycol, with $\xi\le$ 60 \AA\ and for {\it N}-methyl-$\epsilon$-caprolactam ($\xi\le$ 120 \AA) \cite{143}. 
Confinement in porous media, however, introduces curvature effects and is not well defined in size because of the polydispersity in pore sizes and the presence of pore junctions. 
These problems are avoided in thin films, which have so far only been used for polymeric systems. 
In free-standing films, the mean relaxation time was found to be significantly lower than in bulk while the relaxation dynamics followed the same stretched exponential as in bulk \cite{66}. 
In contrast a significant broadening of the distribution of the relaxation times has been observed in films supported by a substrate \cite{132,283}.
There has also been one study of the effect of confinement on the width of distribution of the relaxation times as a function of temperature in a side-chain polymer \cite{132}. 
It was found that, at all temperatures, the average relaxation time is essentially independent of the film thickness, while the width of the distribution of the relaxation times increases as thickness decreases.
The widening of the distribution was however found to be the same over the whole temperature range considered ($T_g-13\;^\circ{\rm C}<T<T_g+25\,^\circ{\rm C}$).
Studies on polymers then have not shown so far any clear trend. 
Because of their chain character, polymers exhibits bulk dynamic processes and a behavior under confinement that is specific to these materials rather than glasses in general. 

To overcome these problems, we have studied the dynamic behavior of low-molecular-weight glass-formers in thin films.  
To our knowledge there is no study reported in literature on the dynamics of low-molecular-weight systems in as simple a confining geometry as thin films. 
We have chosen a hydrogen-bonded glass-former, with a fairly simple chemical structure (see Fig.~\ref{molecules}), from which we can expect relatively simple relaxation processes.
Moreover, small molecules can be obtained with a higher chemical purity and without molecular weight or composition distribution. 
The small size of our compound has however the disadvantage that the characteristic lengths in the system should be smaller than in polymeric systems.
As a consequence, confinement effects will appear at smaller film thicknesses.

In previous measurements performed with x-ray reflectivity \cite{elio1}, we have seen one effect of confinement on a glass former into a thin film: the glass transition significantly widens. 
This was interpreted as originating from an increase of the inhomogeneity of the dynamics in the system.
To obtain a direct evidence for the variation of the inhomogeneity under confinement and study in detail how confinement affects the dynamics of the system, we have examined the reorientation dynamics of our low-molecular-weight glass-former in thin films, starting from an initial state imposed by an external field.
The dynamics has been followed experimentally over a wide range of temperatures around $T_g$, an extended time window and, for confining sizes down to a few molecular units, which is expected to be of the order of $\xi$. 
We have used optical second-harmonic generation technique to follow the re-orientation process. 
Second-order non-linear optical effects, such as second-harmonic generation, are in principle forbidden in centro-symmetric media \cite{136,hein91}. 
It is therefore particularly suitable to follow the re-orientation dynamics between an initial polar state and a final non-polar state. 
\begin{figure}[t!]
\begin{center}
\noindent
{\includegraphics[width=8cm,keepaspectratio]{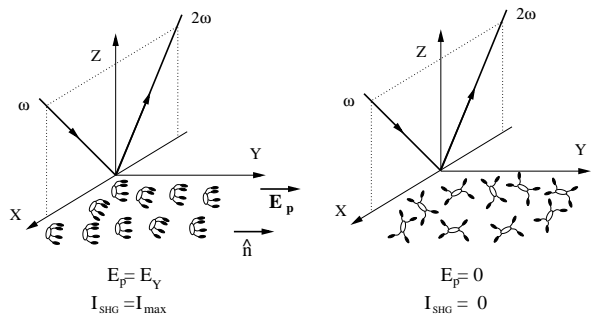}}
\caption{\small Schematic representation of the orientation order of the polar branches in the glass former with applied poling field (left) and after removal of the field and complete relaxation (right). The poling field {\bf E$_Y$} is perpendicular to the optical plane (XZ) defined by the incident and second-harmonic beams.}
\label{shg5}
\end{center}
\end{figure}
The principle of the experiment is the following: the films to be studied are first brought in the liquid phase and a strong DC poling field ${\bf E_p}$, parallel to the film plane, is then switched on.
This field tends to align the dipoles carried by the side-groups along the direction of the field (Fig.~\ref{shg5}).
In this polar state, the sample generates a high second-harmonic (SH) signal. 
At a time $t=0$, we switch the poling field off and follow the reorientation of the dipoles toward an isotropic in-plane distribution of dipoles orientations through the decay of the second-harmonic signal in time. 
Other glassy systems have already been studied with this experimental method \cite{132,146,115,348,349}, most of them being polymeric materials.
In \cite{146} for instance, the measurements probe the reorientation dynamics of a chromophore that is either dissolved in the hosting material or attached to the polymer main chain by a flexible spacer.

In the following, we start by describing our samples and experimental technique (Section \ref{Experimental}). 
We present then our results on the re-orientation dynamics in the glassy films and how it is affected by changes in the size of the system (Section \ref{Results}), and finally discuss these results (Section \ref{disc}).

\section{Experimental}
\label{Experimental}
\subsection{Samples}
\label{procedure}
For the relaxation measurements we have used the material in Fig~\ref{molecules} ($T_g=106/110\,^{\circ}\rm C$) \cite{Akzo,Delft}, that belongs to the polyalcohols family. 
This family is characterized by the presence of hydroxy functional groups that are known to create H-bonded networks. 
In the plot of the relaxation time versus $T_{g}/T$, as originally done by Angell \cite{8,319}, alcohols \cite{300} and polyalcohols \cite{303} are found to occupy the intermediate region between strong and fragile glass-formers.

Thin films were obtained by spin coating from a cyclopentanone solution on fused quartz substrates. 
The thickness (100-1200 \AA) and quality of these films were characterized by x-ray reflectivity \cite{elio1,eliot} and atomic force microscopy \cite{eliot}.
It had already been observed by x-ray reflectivity that films of the same compound  with a thickness between 40 and 100 \AA\ become unstable during the first heating and break up into droplets \cite{315}. 

To be able to apply an electric field parallel to the substrate, the quartz plates were first patterned with two planar electrodes separated by a gap $d$ of 1 mm.
The electrodes were obtained by first depositing a thin (10 nm) layer of chromium, which has a high affinity for silica and then depositing a second thick (100 nm) gold layer.
A high DC electric field was applied between the two electrodes.
For poling the films, we used a constant voltage $V$ of 1.9 kV.
The electric field produced in the gap is perpendicular to the electrode edges and parallel to the substrate surface.
The intensity profile of this field depends on the distance from the edges \cite{346}.
In the middle of the gap the electric field is minimum and equal to $E_p=\frac{2}{\pi}\frac{V}{d}$, which corresponds to 1.2 $\cdot10^6$ V/m.
Around this minimum the field is constant within 1\% in a region of width $d$/5 = 200 $\mu$m in our case.
This width is sufficiently large for our measurements, which are all performed with a projected laser beam area of $\sim$100 $\mu$m in diameter. We assume therefore that the electric field is uniform in the probed surface area.
The degree of polar ordering induced by the poling field is proportional to the applied field. Therefore a higher field is expected to increase the ordering and the initial SH signal before the field is switched off.
Nevertheless we have to limit the applied voltage to 1.9 kV because higher fields cause arching between the two electrodes, which destroys the samples.
The high-voltage power supply limited the current during the transient regime just after switching the electric field on or off to 10 mA.

To eliminate the solvent from the deposited films, samples were first annealed at 45 $^\circ$C for 5 hours. They were then heated above the glass transition to 120 $^{\circ}$C where they were kept for at least 15 minutes.
After switching on the external field the temperature was brought to the desired value and then kept constant during the experiment. All temperature ramps were conducted at 0.25 Kmin$^{-1}$.
To facilitate solvent removal and provide a clean atmosphere, the sample chamber was fluxed continuously with dry nitrogen.

\subsection{Measurements}
The reorientation of the dipoles has been followed measuring the second-harmonic signal after switching off the poling field ${\bf E_p}$. We have measured the second-harmonic signal with an in-coming light and SH signal having both an $s$-polarization ($s$=Y in Fig.~\ref{shg5}, $s$-in-$s$-out) perpendicular to the incident plane (XZ in Fig.~\ref{shg5}).
By releasing of the field at time $t=0$, the field-induced orientational order disappears in the course of time by the randomization of the orientations of the dipoles. 
This leads to a decrease of the SH signal to zero. 
So the reorientation dynamics of the system can be followed by measuring the SH signal $I_{ss}$. 
The square root of the $I_{ss}$ is directly proportional to the degree of polar ordering in the system through the effective susceptibility $\chi_{ss}$ of the film \cite{20}.
The relaxation functions are determined over a range of 11 orders of magnitudes in time from 1 $\mu$s upward.
The ability to characterize reorientation dynamics over a wide range of time scales is crucial to determine accurately the fitting parameters of the relaxation function.  

With the above measurement method, the state from which the relaxation starts is determined by the applied poling field.
For the range of poling fields we have used, $\chi_{ss}(0)$ is proportional to the field strength \cite{132,347,paul2}. In principle, the measured relaxation times can be expected to depend on the degree of order of the starting state of the relaxation, thus on ${\bf E_p}$ \cite{292}. We have not studied this dependence and have used only one value of ${\bf E_p}$.

For the second-harmonic generation measurements we have used a frequency-doubled Q-switched mode-locked Nd-YAG laser (Antares 76, Coherent) as light source with an output wavelength at 532 nm, using a $s$-in-$s$-out polarizations configuration (for more details see \cite{eliot}). 
The repetition rate of the Q-switched pulses was tuned between 50 and 200 Hz depending on the time scale at which the relaxation process was explored. 
A special setup was developed to trigger and delay the switching of the external poling electric field with respect to the optical Q-switched pulses in order to reach time scales for the relaxation process down to 1 $\mu$s \cite{eliot,eliofast}.
The Q-switched pulse energy was at most 1 mJ, which is not high enough to induce molecular reorientation. 
We have also checked that the laser does not influence the dynamics of the system by heating it.
All measurements were averaged over a few thousands repeated collections to increase the signal to noise ratio using a home-developed collection protocol \cite{eliot}.

\subsection{Data analysis}
\label{analysis}
The measured values of the effective susceptibility $\chi_{ss}$ of the film are normalized to the value before the switching off of the poling field ($\chi_{ss}(0)$) and fitted using the phenomenological stretched exponential Kohlrausch-Williams-Watts (KWW) response function \cite{80}:
\begin{equation}
\chi_{\rm NO} (t)=\frac{\chi_{ss}(t)}{\chi_{ss}(0)}=e^{-\left(t/ \tau _W\right)^{\beta_W}}
\label{eq6-1}
\end{equation}
where $\tau _W$ is a characteristic decay time and $0<\beta_W\leq 1$ is the stretch exponent. 
This non-exponential relaxation function is generally associated to a distribution of pure exponential relaxations spread through the system \cite{Donth}.
The system is then considered as a set of relaxators $i$, which can be visualized as nanoscopic regions within the glass-former, each one with a defined time constant $\tau_i$. 
The observed relaxation results from their superposition:
\begin{equation}
\chi_{\rm NO} (t)=\sum_i G(\tau_i) e^{-t/\tau_i}
\label{eq3}
\end{equation}
where the weight of relaxator $i$ is given by $G(\tau_i)$, more often written as a continuous distribution function $G(\tau)$ where $\tau$ is a variable for the relaxation times and is not to be confused with the KWW parameter $\tau _W$.

In the general case when the form of $\chi_{\rm NO} (t)$ is arbitrary, the two functions $\chi_{\rm NO} (t)$ and $G(\tau)$ are uniquely derived from one another through an operation involving a Laplace transform \cite{Lindsey}.
In the case when $G(\tau)$ follows the KWW law, $G(\tau)$ is directly determined by the two parameters $\tau_W$ and $\beta_W$. 
$G(\tau)$ can then be written in the form of a definite integral or a series, but there is no analytical solution except for the cases $\beta_W=1/2$ and $\beta_W=1$.

The distribution function $G(\tau)$ is numerically computed using a recursive computer algorithm developed by Emri and Tschoegl \cite{Emri}, and often applied to interpret SHG decays \cite{200,132,35}.
The maximum of $G(\tau)$ corresponds approximately to $\tau_W$ and its breadth is related to $0<\beta_W\leq 1$. 
The smaller $\beta_W$ the wider the distribution, the special case $\beta_W=1$ corresponding to a single-exponential relaxation and a delta function $G(\tau)$.

\section{Results}
\label{Results}
We have examined the dependence of the reorientation dynamics as a function of two parameters: temperature and film thickness. In the following, we first present the results concerning the temperature dependence of the dynamics in a thick film (as-spun thickness $h_0$=116 nm). 
\begin{table}
\caption{Values of the characteristic time $\tau_W$ and stretch exponent $\beta_W$ obtained by fitting the experimental relaxation with Eq.~(\ref{eq6-1}), and average relaxation $\langle\tau\rangle$ time calculated with Eq.~(\ref{eq6-2}), for the different temperatures and thicknesses considered.}
\begin{ruledtabular}
\begin{tabular}{lrrcr}
$h(0)$&T ($^\circ$C)& $\tau$ (sec)&$\beta$&$\langle\tau\rangle$ (sec)\\
\hline
116 nm&60 & 1.46 $\cdot10^6$ & 0.26 & 29 $\cdot10^6$\\
&80 & 36 $\cdot10^3$   & 0.32 & 250 $\cdot10^3$\\
&90 & 16 $\cdot10^3$   & 0.28 & 220 $\cdot10^3$\\
&100 & 1.1 $\cdot10^3$ & 0.34 & 6.2 $\cdot10^3$\\
&120 & 160 & 0.31 & 1.24 $\cdot10^3$\\
70 nm & 100 & 2.5 $\cdot10^3$  & 0.31 & 20.5 $\cdot10^3$\\
41 nm & 100 & 6.4 $\cdot10^3$ & 0.30 & 62 $\cdot10^3$\\
25 nm & 100 & 1.5 $\cdot10^3$  & 0.27 & 25 $\cdot10^3$\\
17 nm&60 & 1.5 $\cdot10^6$ & 0.24 & 43 $\cdot10^6$\\
&90 & 170 $\cdot10^3$ & 0.21 & 15 $\cdot10^6$\\
&100 & 38 $\cdot10^3$ & 0.20 & 3.7 $\cdot10^6$\\
&110 & 2.9 $\cdot10^3$ & 0.24 & 93 $\cdot10^3$\\
&120 & 160 & 0.32 & 1.1 $\cdot10^3$\\
&130 & 39 & 0.56 & 65\\
10 nm&60 & 6 $\cdot10^6$ & 0.26 & 110 $\cdot10^6$\\
&90 & 5.9 $\cdot10^3$ & 0.22 & 320 $\cdot10^3$\\
&100 & 2.8 $\cdot10^3$ & 0.21 & 190 $\cdot10^3$\\
&110 & 610 & 0.28 & 7.8 $\cdot10^3$\\
\end{tabular}
\end{ruledtabular}
\label{tab}
\end{table}
From the thermal expansivity measurements we have presented in \cite{elio1}, we expect a bulk-like
behavior for the molecular dynamics of such a thick film. We can therefore compare our dynamic measurements with the expected behavior around the glass transition temperature of a bulk glassy system.
Next we examine how the dynamics at different temperatures is affected as the film thickness decreases.
From the broadening of the glass transition we have reported in \cite{elio1}, which is a sign of an increased inhomogeneity of the film, we expect a broadening of the distribution of the relaxation times as the film thickness decreases.

\subsection{Temperature-dependence of relaxation dynamics in thick films}
\label{temper}
\begin{figure}[t!]
\begin{center}
\resizebox{8.0cm}{!}
{\includegraphics{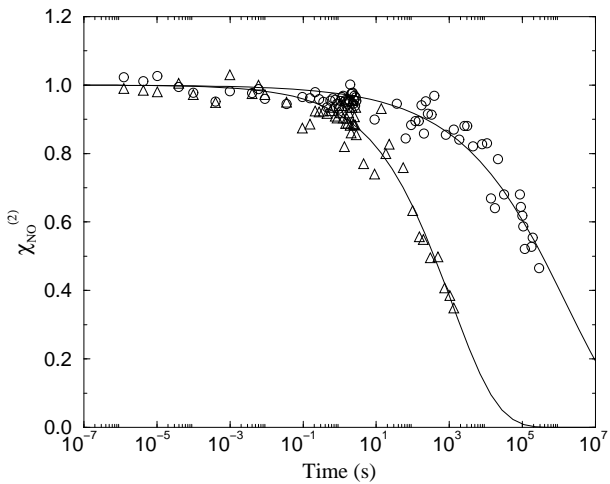}}
\caption{\small Normalised second-order susceptibility vs.~time for a 116 nm-thick film (as-spun thickness) at ($\circ$) 60 $^\circ$C ($T_g$-46) and ($\vartriangle$) 100 $^\circ$C ($T_g$-6). Solid lines represent the best least-square fit using Eq.~(\ref{eq6-1}).}
\label{1200}
\end{center}
\end{figure}

In Fig.~\ref{1200} we plot the time dependence of the effective non-linear susceptibility $\chi_{ss}$ normalized to its value before the poling field is switched off at time $t=0$, for a 116 nm-thick film.
We show for clarity the results only for the temperatures just below (100 $^{\circ}$C) the $nominal$ bulk glass transition temperature determined by differential scanning calorimetry (DSC) ($T_g$=106 $^{\circ}$C) and deep into the glass phase (60 $^{\circ}$C).
From the observed decays it is clear that any contribution to $\chi_{ss}$ from electric-field-induced second-harmonic (EFISH) generation \cite{eliofast,paul,paul2} can be neglected. 
Indeed, for all temperatures, the SH signal does not decrease for several decades in time. 
A possible contribution of EFISH would result in an immediate jump of the signal to a lower value, as fast as the poling field ${\bf E_p}$ drops to zero ($\sim100$ ns). 
We conclude then, that the observed decays are associated solely to reorientation dynamics of the non-linearly polarizable units of the molecules.
\begin{figure}[t!]
\begin{center}
\noindent
\resizebox{8.0cm}{!}
{\includegraphics{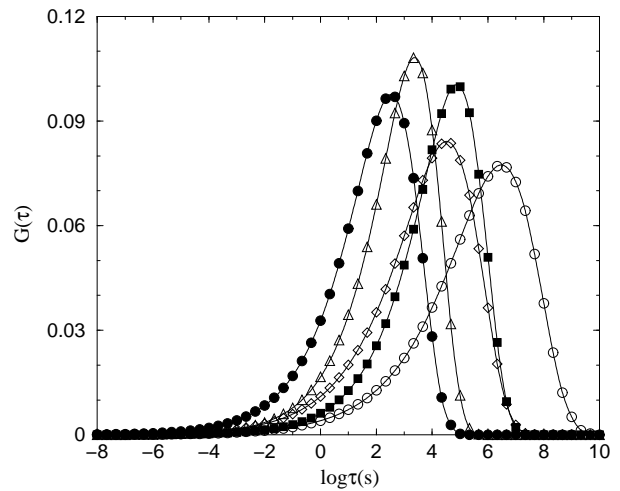}}
\caption{\small Distribution of relaxation times calculated for a 116 nm-thick film from the parameters given in Table \ref{tab}: 60 $^\circ$C ($\circ$), 80 $^\circ$C ($\blacksquare$), 90 $^\circ$C ($\diamond$), 100 $^\circ$C ($\vartriangle$) and 120 $^\circ$C ($\bullet$). The solid lines are only a guide to the eye.}
\label{1200tau}
\end{center}
\end{figure}

Data are fitted with the stretched-exponential KWW law given by Eq.~(\ref{eq6-1}).
The characteristic time $\tau_W$ and the stretch exponent $\beta_W$ obtained at the different temperatures are given in Table \ref{tab}.
The values of $\beta_W$ exhibit little temperature dependence. 
As they are significantly different from 1, it is obvious that the relaxation process is highly non-exponential.

Another way to characterize the dynamic behavior of the system is to calculate an average reorientation time $\langle\tau\rangle$ as is generally done for polymer systems from dynamic optical or dielectric measurements \cite{330,73,329,307,115}:
\begin{equation}
\langle\tau\rangle=\int^\infty_0\exp\left[-\left(\frac{t}{\tau_W}\right)^{\beta_W}\right]{\rm d}t=\frac{\tau_W \Gamma\left(1/\beta_W\right)}{\beta_W}
\label{eq6-2}
\end{equation}
where $\Gamma$ is the gamma function. 
For $\beta_W=1$ (single exponential), $\langle\tau\rangle$ is equal to $\tau_W$. 
The results for the 116 nm-thick film are given in Table \ref{tab}. 
For all temperatures $\langle\tau\rangle$ is rather large.
Even for the highest temperature ($T=120 ^\circ$C$ =T_g+14 ^\circ$C) the average relaxation time lies above the value generally associated to the macroscopic glass transition $\langle\tau\rangle_{Tg}=100$ s \cite{80}.

From the KWW fitting parameters $\tau_W$ and $\beta_W$ we have computed the distribution functions of relaxation times $G(\tau)$ as explained in Section \ref{analysis} (Fig.~\ref{1200tau}). 
The measured relaxation decays at low temperatures do not decay to 0 at the largest experimental time (Fig.~\ref{1200}).
Calculating distribution of relaxation times using then high characteristic time $\tau_W$ evaluated with the KWW fitting law (Eq.~(\ref{eq3})) introduces weights $G(\tau_i)$ for times $\tau_i$ longer than the measured ones.
Thus the contributions to the distribution function at times $\log(\tau)>6$ have to be taken as purely a result of the computation but not directly related to the experimental measurements.
The calculations of $G(\tau)$ confirm the trend indicated by Table \ref{tab} for the average relaxation time and the width of the distribution.
For the considered film thickness, the width of the distribution does not show a particular trend at temperatures larger than 80 $^\circ$C. 
\begin{figure}[t!]
\begin{center}
\noindent
\resizebox{8.0cm}{!}
{\includegraphics{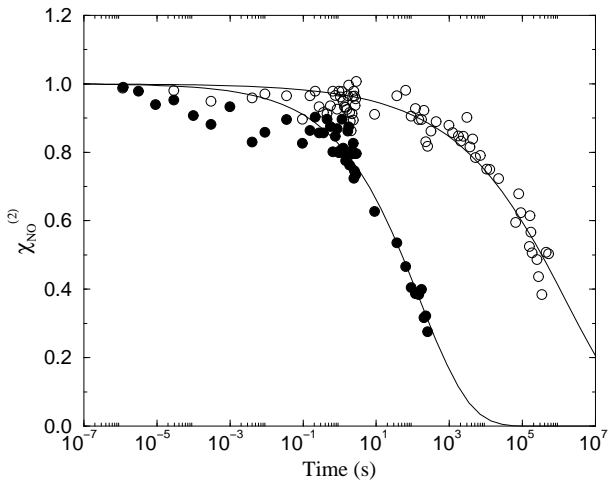}}
\caption{\small $\chi^{(2)}_{\rm NO}$ vs.~time for a 17 nm-thick film (as-spun)  at ($\circ$) 60 $^\circ$C ($T_g$-46) and ($\bullet$) 120 $^\circ$C ($T_g$+14). Solid lines represent the best least-square fit using 
Eq.~(\ref{eq6-1}).}
\label{17}
\end{center}
\end{figure}
\begin{figure}[t!]
\begin{center}
\noindent
\resizebox{8.0cm}{!}
{\includegraphics{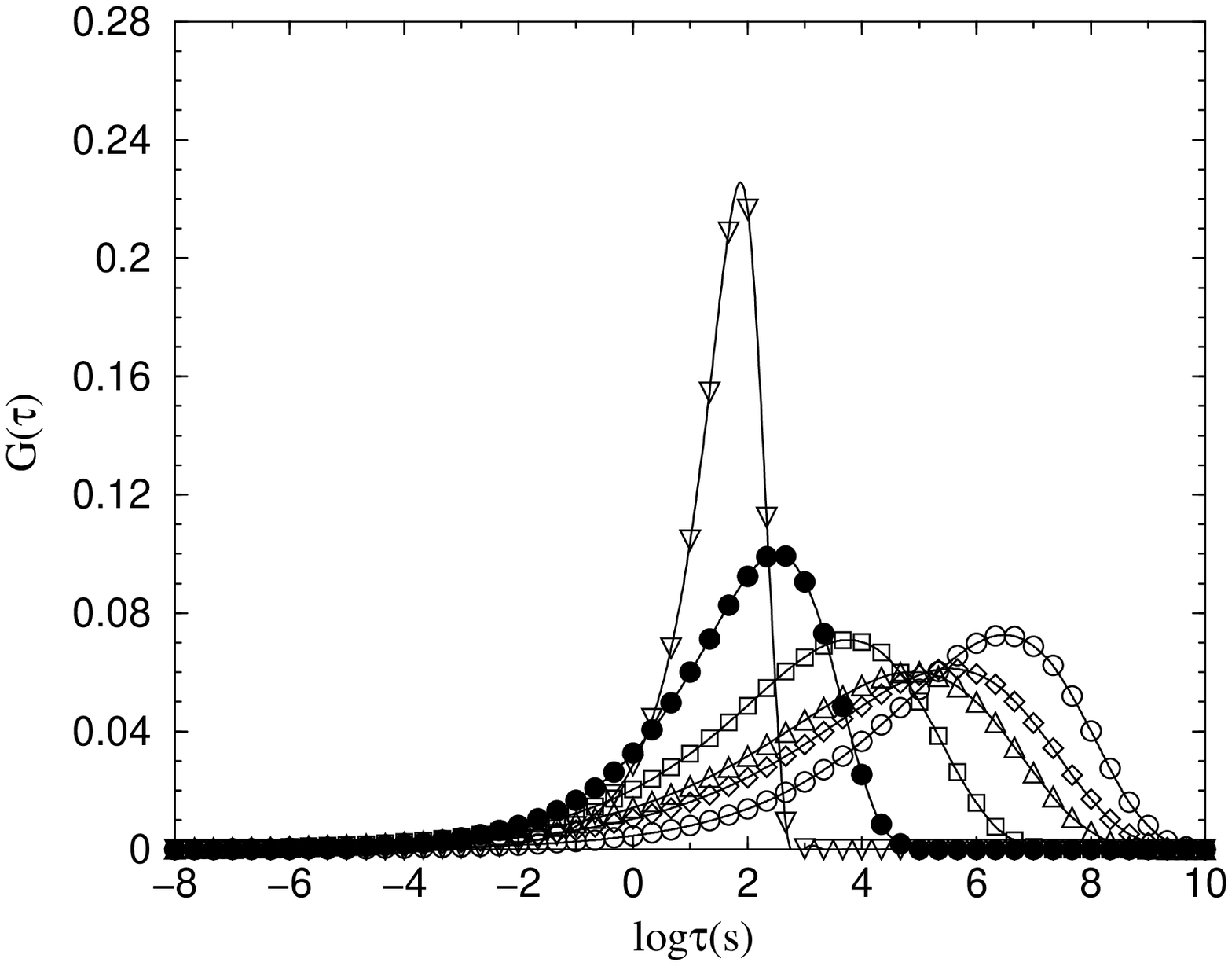}}
\caption{\small Distribution of relaxation times  for a 17 nm-thick film
from the parameters in Table \ref{tab}: 60 $^\circ$C ($\circ$), 90 $^\circ$C
($\diamond$), 100 $^\circ$C ($\vartriangle$), 110 $^\circ$C ($\square$), 120
$^\circ$C ($\bullet$) and 130 $^\circ$C ($\triangledown$). The solid lines are a
guide to the eye.}
\label{17tau}
\end{center}
\end{figure}
Only at the lowest temperature measured (60 $^\circ$C) a slight widening of the distribution is visible.
When the temperature increases, the peaks of the distribution move from long to short times by almost four orders of magnitude in time in the explored temperature window.
Note that the characteristic time $\tau_W$ corresponds approximately to the maximum of the distribution, while the average calculated with Eq.~(\ref{eq6-2}) is equal to the average calculated from the distribution $G(\tau)$. This confirms the consistency of our data analysis.

The wide distributions of relaxation times shown in Fig.~\ref{1200tau} confirms the fact that to make an accurate evaluation of the distribution $G(\tau)$, experimental relaxation decays covering many orders of magnitude in time are needed. Indeed, the relaxation time distributions cover 14 orders of magnitude in time ($10^{-4}-10^{10}$ s) for the studied range of temperatures.

\subsection{Thickness-dependence of the relaxation dynamics}
To investigate the effect of confinement we carried the same relaxation experiment as presented in Sec.~\ref{temper} for films with a lower thickness.
In Fig.~\ref{17} we show the signal decay measured at 120 $^\circ$C and 60 $^\circ$C for a 17 nm-thick film (as-spun thickness).
\label{size}

The decay measured at 120 $^\circ$C presents an interesting feature: the SH signal exhibits a first decay in the short time domain.
The use of the KWW equation to fit the data appears to be only a first approximation.
The data are not properly described in the ms time window and a general double stretched exponential law would be more suitable.
It seems that, under our experimental conditions, it is possible to discriminate different relaxation processes.
Measuring relaxation dynamics around and below the bulk glass transition means in principle probing the $\alpha$-relaxation process.
Nevertheless, at high temperatures, the $\alpha$-process does not necessary screen all other faster processes.
These modes observed at high temperature can have different origins. 
One possible interpretation is to attribute this mode to the fast $\beta$-process, related to a local mobility.

However, given the small contribution of possible fast processes (observed only at high $T$), we fitted in first approximation all decays with a single stretched exponential function using Eq.~(\ref{eq6-1}).
Fitting with a double stretched exponential the data at high $T$ hardly changes the parameters for the $\alpha$-process.
The resulting fitting parameters $\tau_W$ and $\beta_W$ together with the calculated average relaxation time $\langle\tau\rangle$ are reported in Table \ref{tab}.
The corresponding distribution of relaxation times $G(\tau)$ are plotted in Fig.~\ref{17tau}.

The distribution obtained in the glass phase are slightly wider than the ones obtained with the 116 nm-thick film. They also show a clear increase of the exponential character (increase of $\beta_W$) for the highest temperature measured, which results in a sharper distribution of relaxation times for $T$=130 $^\circ$C in Fig.~\ref{17tau}.
Note that for this temperature the distribution function moves towards shorter times and $\langle\tau\rangle<100$ s.

To increase the effect of confinement, we repeated the above experiments with an even thinner film. 
The only difference with the thicker films was that the annealing temperature was 110 $^\circ$C instead of 120 $^\circ$C, due to the instability of the film at higher temperatures.

\begin{figure}[t!]
\begin{center}
\noindent
\resizebox{8.0cm}{!}
{\includegraphics{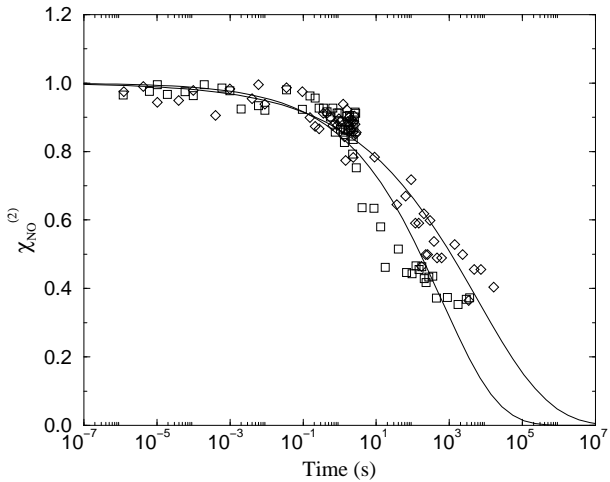}}
\caption{\small $\chi^{(2)}_{\rm NO}$ vs.~time for a 10 nm-thick film
(as-spun thickness) at  90 $^\circ$C ($T_g$-16) ($\diamond$) and $^\circ$C
($T_g$+4) ($\square$) 110. Solid lines represent the best least-square fit using
Eq.~(\ref{eq6-1}).}
\label{10}
\end{center}
\end{figure}

\begin{figure}[t!]
\begin{center}
\noindent
\resizebox{8.0cm}{!}
{\includegraphics{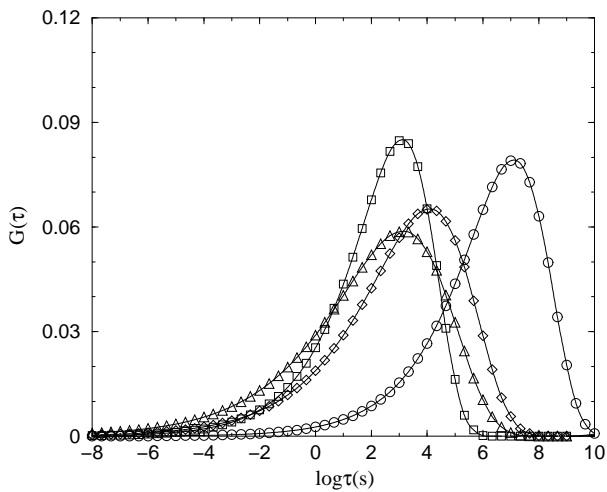}}
\caption{\small Distribution of relaxation times for a 10 nm-thick film from
the parameters in Table \ref{tab}: 60 $^\circ$C ($\circ$), 90 $^\circ$C
($\diamond$), 100 $^\circ$C ($\vartriangle$) and 110 $^\circ$C ($\square$). The
solid lines are a guide to the eye.}
\label{10tau}
\end{center}
\end{figure}

Fig.~\ref{10} shows the relaxation for the 10 nm-thick film at 90 $^\circ$C and 110 $^\circ$C. For this film thickness experimental data are in general of lower quality with a higher scattering of the points. This is mainly due to the low value of the SH signal intensity in the oriented state at $t<0$. Analysing ultra-thin films reduces the number of molecules probed by the input laser beam, reducing the sensitivity of the technique.
The fitting parameters of the SH decays for four different temperatures are presented in Table \ref{tab}.
The calculated distribution of relaxation times are shown in Fig.~\ref{10tau}.

As temperature increases, the peak in the distribution shifts uniformly towards shorter times till 100 $^\circ$C.
As for the 17 nm-thick film, $\beta_W$ first decreases and then increases slightly as the temperature decreases.
In both cases, the minimum in $\beta_W$ is at around 100 $^\circ$C ($T_g$-6 $^\circ$C).

At this temperature we have also collected the relaxation decay for 25 nm-thick, 41 nm-thick and 70 nm-thick films. The distribution of relaxation times $G(\tau)$ of all the measured films at 100 $^\circ$C are plotted in Fig.~\ref{cnfbetatau}.

\section{Discussion and conclusion}
\label{disc}
We summarise our results on the dynamics of thin films by plotting the temperature dependence of the average relaxation time $\langle\tau\rangle$ (Fig.~\ref{alltau}) and the stretch exponent $\beta_W$ (Fig.~\ref{allbeta}) for all the films we have measured.

The first important issue arising from these measurements is whether the measured reorientation dynamics of the non-linearly polarisable groups is coupled with any glassy dynamics of the molecules as a whole.
An element of answer can be found in the values of $\beta_W$ for the thickest film exhibiting a bulk-like behaviour. The values are rather scattered but remain in the interval 0.25-0.35.
If we take for the value at $T_g$ the average of the measured values, we obtain $\beta_W(T_g)\sim0.30$, which is close to the value for bulk poly(propylene glycol).
This can be expected from the fact that our compound carries four hydroxy groups per molecules and should belong to the most fragile hydrogen-bonded glass-formers.
Another element of answer can be found in the Arrhenius plot of $\langle\tau\rangle$ as a function of the inverse temperature (Fig.~\ref{alltau}) where $\langle\tau\rangle$ increases exponentially with $1/T$ as expected for a glassy material.

We thus conclude that the reorientation dynamics we observe, corresponds to the one expected for a glass-former.
The reorientation of the non-linearly polarisable groups is probably strongly coupled  to that of the rest of the molecules, since the flexible spacer between the central rigid group and the polarisable side-groups contains only three atoms (see Fig~\ref{molecules}).
The fact that the reorientation dynamics of side-groups is coupled to the $\alpha$-relaxation process has also been demonstrated in a polymeric system with a flexible spacer of four atoms \cite{200}.
The same coupling has also been demonstrated for chromophores dissolved in a polymer without any covalent bonding with the polymer chains \cite{115}.
However this could be due to strong steric interactions between the chromophores and the chains, which are less likely to take place in our low-molecular weight glass-former.

We also observe however deviations of the dynamics behavior from the expectations on two points.
One is that even at temperatures above the bulk glass transition temperature $T_{g}$ (as measured by DSC), the average relaxation time $\langle\tau\rangle$ is longer than the one corresponding to the standard definition of $T_{g}$: $\langle\tau\rangle_{T_g}=100$ s.
In \cite{elio1} we saw that there is no evident variation of $T_{g}$ under confinement.
So the high values of $\langle\tau\rangle$ cannot be explained by an increase of ${T_g}$.
The origin of the long relaxation times might be related to the fact that the non-polarisable side-groups are polar mesogenic units. Ordinary non-glassy liquid crystals based on the same cyanobiphenyl units exhibit also a very slow reorientation dynamics in ultra-thin films (thickness below 10 nm) \cite{paul,pault,paul3}. The reason for this slow dynamics is not clear.
One possible explanation could be the electrostatic correlation between the dipoles carried by the cyanobiphenyl groups \cite{paul2}, although no direct evidence for such a correlated dynamics has been found yet.
\begin{figure}[t!]
\begin{center}
\noindent
\resizebox{8.0cm}{!}
{\includegraphics{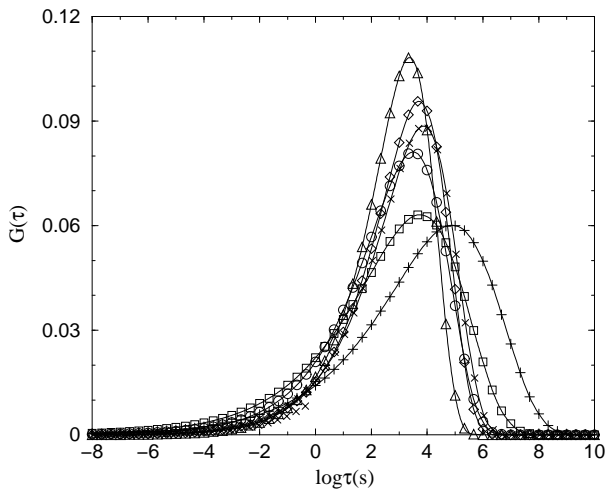}}
\caption{\small Distribution of relaxation times at 100 $^\circ$C ($T_g$-6) 
for a 10 nm ($\square$), 17 nm ($\rm{+}$), 25 nm ($\circ$), 41 nm ($\times$), 70
nm ($\diamond$) and a 116 nm ($\triangle$) thick films. The solid lines are a
guide to the eye.}
\label{cnfbetatau}
\end{center}
\end{figure}

The second unexpected feature in the observed dynamics (in particular in thick films) is the fact that the time dependence of the observed relaxations remains strongly stretched at temperatures above the bulk glass transition temperature.
We could expect an increase of the stretch exponent $\beta_W$ indicating a narrowing of the distribution as temperature increases. The fact that we do not observe this narrowing might simply be due to the limited range of temperature that we can explore above $T_g$. When we have managed to performed experiments at slightly high temperatures (130 $^\circ$C with the 17 nm-thick film), we indeed observed an increase of $\beta_W$.
Similar results were found on polymer systems \cite{35,115,100}, where $\beta_W$ does not vary significantly in the glassy phase, while for temperature above $T_g$, $\beta_W$ was increasing with temperature.

Our experimental results allow a discussion on the main purpose of our study, namely the effect of confinement on the dynamics of glass-formers.
Let us start with the average relaxation time $\langle\tau\rangle$ (Fig.~\ref{alltau}).
At high and low temperature, $\langle\tau\rangle$ hardly depends on the film thickness.
At intermediate temperatures around $T_g$, there is a large scatter without a specific variation with film thickness.
This is likely due to the limited relevance of $\langle\tau\rangle$ when the distribution of relaxation times $G(\tau)$ becomes very wide.
Variations in the actual shape of the distribution can have a strong influence on $\langle\tau\rangle$ while the location of the maximum of $G(\tau)$ hardly moves.
This can be seen in the plot of the distribution of relaxation times for different film thickness at $T=$ 100 $^\circ$C (Fig.~\ref{cnfbetatau}).
Except for the thickness of 17 nm, maxima are hardly shifted with respect to one another (less than one order of magnitude), but the asymmetric widening of $G(\tau)$ for the thin films leads to differences in $\langle\tau\rangle$ of two orders of magnitude.
So the average relaxation time $\langle\tau\rangle$ is not the most relevant parameter to describe the variation of the distribution of relaxation times with film thickness.
One should rather look at the width of the distribution, which is related to the stretch exponent $\beta_W$.
The summary of the results in Fig.~\ref{allbeta} show that we can distinguish three temperature regions exhibiting different behaviours:
\begin{itemize}
\item {\bf high temperatures} 
($T\gtrsim 120\;^\circ$ C): although the number of measurements we could perform in this range is small, it is clear that thickness has little influence on the dynamics. This can be understood from the fact that at 120 $^\circ$C, in the liquid phase, the dynamics can only be little affected by size effects. Eventually the dynamics will be modified at small thickness (below 10 nm) by the presence of the free surface and film-substrate interface.
The effect is quite likely not large enough to be observed with our experiments.
\begin{figure}[t!]
\begin{center}
\noindent
\resizebox{8.0cm}{!}
{\includegraphics{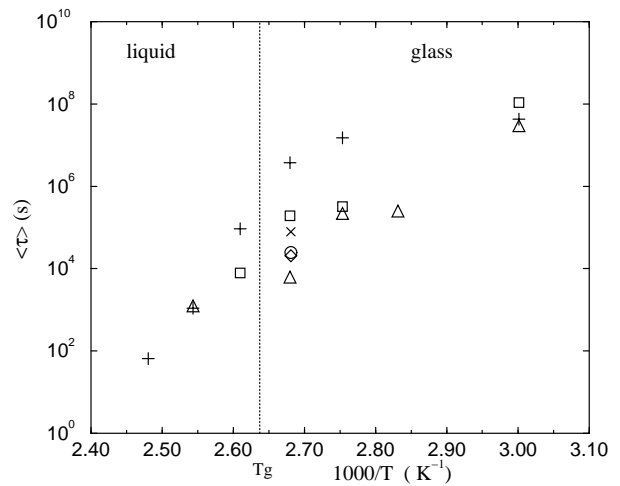}}
\caption{\small Average relaxation time $\langle\tau\rangle$ vs.~inverse
temperature for 10 nm ($\square$), 17 nm ($\rm{+}$), 25nm ($\circ$), 41 nm
($\times$), 70 nm ($\diamond$) and  116 nm ($\triangle$) thick films.}
\label{alltau}
\end{center}
\end{figure}
\begin{figure}[t!]
\begin{center}
\noindent
\resizebox{8.0cm}{!}
{\includegraphics{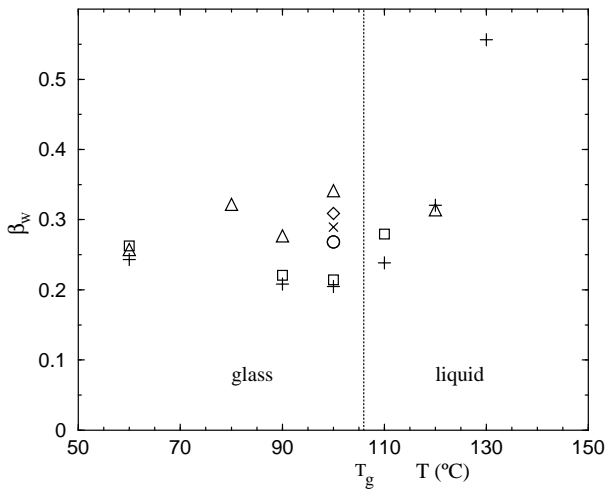}}
\caption{\small Stretch exponent $\beta_W$ vs.~temperature for 10 nm
($\square$), 17 nm ($\rm{+}$), 25nm ($\circ$), 41 nm ($\times$), 70 nm
($\diamond$) and  116 nm ($\triangle$) thick films.}
\label{allbeta}
\end{center}
\end{figure}
\item {\bf intermediate temperatures} 
($70 \;^\circ{\rm C}\lesssim T \lesssim 120\;^\circ$ C): in this temperature range, there is an influence of the film thickness on the dynamics.
 First of all the stretch exponent $\beta_W$ of the thinnest films (10 nm and 17 nm) significantly decreases as temperature decreases and reaches a minimum at around 100 $^\circ$C and then slightly increases again (Fig.~\ref{allbeta}).
The decrease of $\beta_W$ indicates a widening of the distribution of the relaxation times.
This widening is also observed for intermediate thicknesses at $T=$ 100 $^\circ$C, but the effect is smaller and decreases as the film thickness increases.
Since this widening is not observed in the thickest film (116 nm-thick) we can conclude that it is due to an increase of the inhomogeneity of the films under confinement. This is in agreement with our observation of a broadening of the glass transition as the film thickness decreases.
Note also that the temperature range in which $\beta_W$ is decreased in the thinnest films, with respect to the value in thick films, corresponds to the temperature range in which the glass transition takes place in the thinnest films studied with x-ray reflectivity \cite{elio1}.
This is indeed the temperature range in which one expects the inhomogeneity to be the largest, with part of the film in the liquid state and the rest in the glassy state.
Since the average relaxation time varies very rapidly in the vicinity of the glass transition, the  difference between the dynamics in the different parts of the system is then the largest.

\item {\bf low temperatures}
($T\lesssim60\;^\circ$C): the difference in dynamics between films at different thicknesses diminishes as temperature decreases below 100 $^\circ$C and becomes unmeasurable at approximately 60 $^\circ$C. Although we have not performed measurements at lower temperatures because the relaxation of the system becomes extremely long, we can expect that this independence on temperature also occurs below 60 $^\circ$C.

Following the argument in the discussion of the intermediate temperatures, we can say that 60 $^\circ$C is the temperature at which the glass transition is completed in the whole film (for the thinnest films we have measured).
The inhomogeneity of the dynamics induced by confinement becomes then equal to or smaller than the intrinsic inhomogeneity of the dynamics in bulk and thick films.

Our finding that $\beta_W$ becomes independent of film thickness at low temperatures is is contrast with the results of Hall {\it et al.}~\cite{132} on polymer films. They found little influence of temperature on the difference in value of $\beta_W$ for thin and thick films. This might however be due to the fact that they only performed measurements for temperatures above
$T_g-10$ $^\circ$C, while we measured down to $T_g-46$ $^\circ$C. 
\end{itemize}

In the above discussion, we have not considered the role of the cooperative length $\xi$ and its temperature dependence. From the model presented in \cite{113}, we know that the existence of $\xi$ introduces an inhomogeneity of the system for film thicknesses close to $\xi$.
The original theory of Adam and Gibbs of cooperative motion \cite{69} predicts a divergence of $\xi$ at a temperature below $T_g$. 
This implies that $\xi$ should increase as temperature decreases. 
This means that as temperature decreases, the effect of confinement on the heterogeneity of the dynamics should start appearing at larger thicknesses and, for a given thickness, the inhomogeneity should increase as temperature decreases.
This is however not what we have observed. One explanation for this is that $\xi$ does not increases as temperature decreases, or at least not much. 
This has actually been observed in colloidal suspension using confocal microscopy \cite{335,341}. 
In such a system, the glass transition is obtained by increasing the density of particles. 
The size of fast relaxing clusters (which can be compared to cooperatively rearranging regions) was found to remain constant when the density increases beyond the glass transition. 
Unfortunately, no data is available on the temperature dependence of $\xi$ below $T_g$ in molecular systems.

In the simple model we presented in \cite{267}, making the cooperative length $\xi$ independent of temperature, makes also the relaxation time in bulk and the degree of inhomogeneity in the film independent of temperature. 
We could introduce a variation of relaxation time in bulk with temperature in a phenomenological way by imposing the appropriate temperature dependence of the energy barrier per molecule that needs to be overcome for allowing for a relaxation process \cite{267}.
There is however no obvious way of obtaining a dependence of the distribution of relaxation times in the films as a function of temperature in agreement with our experimental results.
So the simple model based on the idea that motion is cooperative over a certain length scale is not able to explain all the dynamic feature of thin glassy films, although it explains qualitatively some of them.

In conclusion, we have investigated how the distribution of relaxation times in a low-molecular-weight glass-former is affected by confining the sample in a thin film.
Our main finding is that the effect of confinement on the dynamics of the system is the largest around the glass transition temperature in bulk and that this effect vanishes as temperature is further lowered deep in the glass phase. 
\acknowledgments This research was supported by the Council for Chemical Sciences of the Netherlands Organization for Scientific Research (CW-NWO).

\bibliography{literature}

\begin{thebibliography}{56}
\bibitem[{Del()}]{Delft}
\bibinfo{note}{A second batch was synthesised by W.~Jager and P.~Kouwer in the
  Polymer Materials and Engineering group at the University of Delft.}
\end{thebibliography}

\end{document}